\documentclass[review]{elsarticle}

\usepackage{ulem}
\usepackage{graphics}
\usepackage{amssymb}
\usepackage{multirow}
\usepackage{booktabs}
\usepackage{threeparttable}
\usepackage{gensymb}

\usepackage{amssymb}
\usepackage{color}

\definecolor{mygreen}{RGB}{0, 102, 0}

\journal{Journal of Magnetism and Magnetic Materials}
\begin{document}
%%%%%%%%%%%%%%%%

%%%%%%%%%%%%%%%%%%%
\begin{frontmatter}
%%%%%%%%%%%%%%%%%%%



%%%%%%%%%%%%%%%%%%%%%%%%%%%%%%%%%%%%%%%%%%%%%%%%%%%%%%%%%%%%%%%%%%%%%%%%%%%%
\title{Structural and Magnetic Properties of  Fe-Al alloys: an \textit{Ab initio} Studies}
%%%%%%%%%%%%%%%%%%%%%%%%%%%%%%%%%%%%%%%%%%%%%%%%%%%%%%%%%%%%%%%%%%%%%%%%%%%%


%\def\correspondingauthor{\footnote{Corresponding author:}}

\author{Mikhail~A.~Zagrebin }
\author{Mariya~V.~Matyunina\corref{cor1}}
\cortext[cor1]{Corresponding author}
              \ead{matunins.fam@mail.ru}
\author{Alexey~B.~Koshkin\corref*{}}
\author{Vladimir~V.~Sokolovskiy}
\author{Vasiliy~D.~Buchelnikov}  

\address{Chelyabinsk State University, 454001 Chelyabinsk, Russia}

%\address[label2]{National University of Science and Technology ''MISIS'', 119991 Moscow, Russia}

%%%%%%%%%%%%%%%%%%%%%%%%%%%%%%%%%%%%%%%%%%%%%%%%%%%%%%%%%%%%%%%%%%%%%%%%%%%%
\begin{abstract}
%%%%%%%%%%%%%%%%%%%%%%%%%%%%%%%%%%%%%%%%%%%%%%%%%%%%%%%%%%%%%%%%%%%%%%%%%%%%
%
In the framework of density functional theory, the structural and magnetic properties of Fe$_{100-x}$Al$_x$ alloys (${5 \leq x \leq 25}$~at.\%) with the different structural order are investigated.
Using the Korringa-Kohn-Rostoker Green's function method with a coherent potential approximation, the equilibrium lattice parameters, ground-state energy, and shear moduli for D0$_3$, B2, and A2 structures are calculated. For all structures,
the optimized lattice constant increases while the shear modulus demonstrates a decreasing behavior with increasing Al content.
The tetragonal magnetostriction constants are estimated by the torque method.
The A2 and B2 structures provide a positive contribution to the tetragonal magnetostriction.
With the help of Monte Carlo simulations of the Heisenberg model, the Curie temperatures are obtained in a wide concentration range.
%The theoretical phase diagram of the magnetic and structural phase transitions for Fe$_{100-x}$Al$_x$ is plotted. 

%%%%%%%%%%%%%%%%%%%%%%%%%%%%%%%%%%%%%%%%%%%%%%%%%%%%%%%%%%%%%%%%%%%%%%%%%%%%
\end{abstract}
%%%%%%%%%%%%%%%%%%%%%%%%%%%%%%%%%%%%%%%%%%%%%%%%%%%%%%%%%%%%%%%%%%%%%%%%%%%%

%%%%%%%%%%%%%%%%%%%%%%%%%%%%%%%%%%%%%%%%%%%%%%%%%%%%%%%%%%%%%%%%%%%%%%%%%%%%
%%%%%%%%%%%%%

\end{frontmatter}



\section{Introduction}

\begin{figure}[!htb]
\centering
\includegraphics[height=6cm,clip]{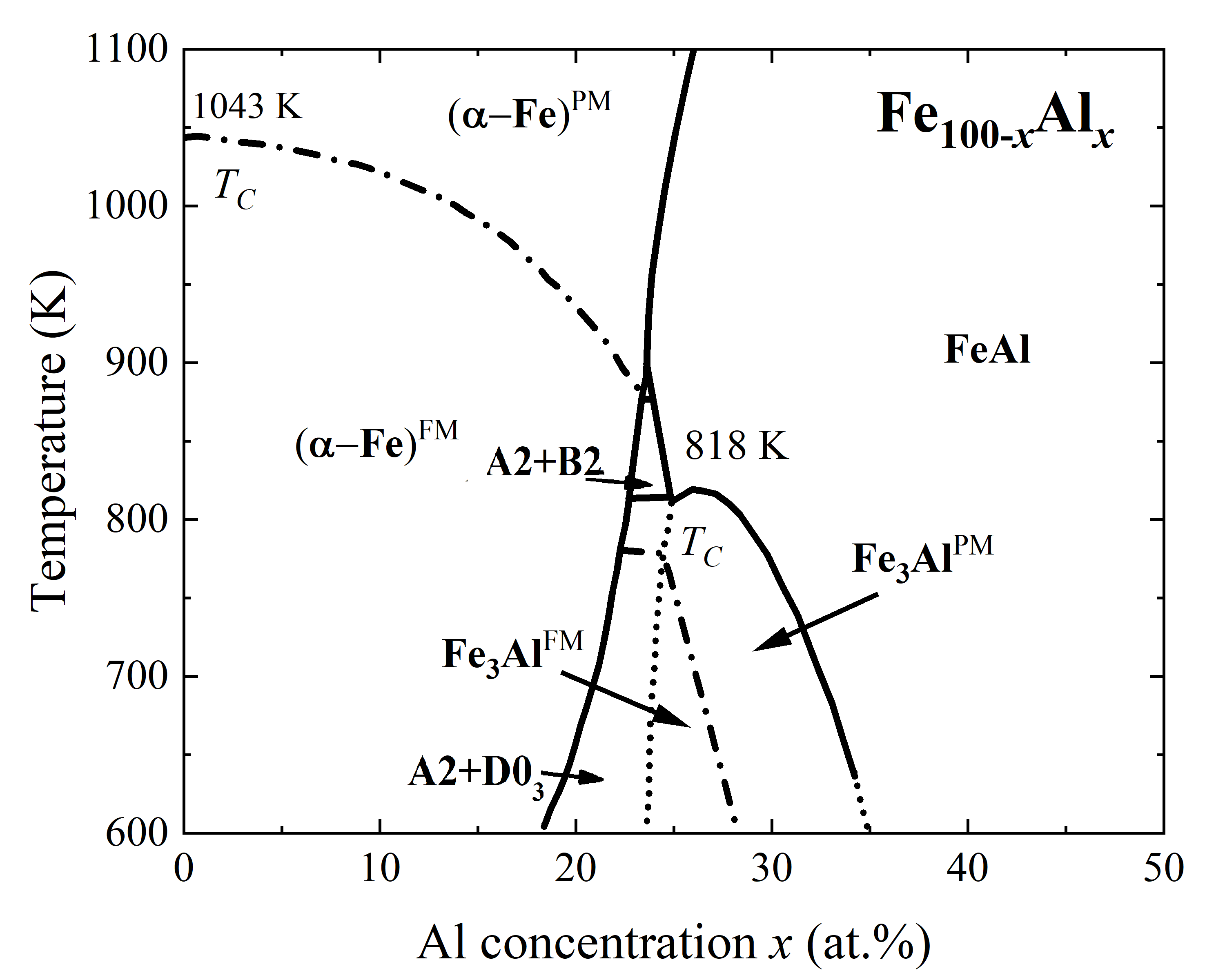}
\hfil
\caption{Experimental phase diagram of Fe$_{100-x}$Al$_x$ (${0\leq x \leq 50}$) adapted from~\cite{Stein}. The dash-dotted lines mark the transitions from ferromagnetic~(FM) to paramagnetic~(PM) $\alpha-$Fe and Fe$_3$Al. Dotted line between about 23 and 24 at.\% Al indicates ($\alpha-$Fe + Fe$_3$Al)/Fe$_3$Al boundary. }
\label{fig_0}
\end{figure}

Fe-Al alloys are attractive materials that exhibit a good oxidation and sulfidation resistance, excellent resistance to abrasive wear and erosion, high strength, relatively low density and high magnetic permeability~\cite{Knibloe1993,Eggersmann2000,Ikeda2001}.
Due to these unique properties, Fe-Al alloys are promising as the high-temperature functional materials for magnetic and diffusion barrier applications,  and interconnections in microelectronics corrosion stability.
{The functional properties of ordered alloys strongly depend on their phase structure.
According to the experimental phase diagram (see Fig.~\ref{fig_0}), three bcc structures (A2, B2 and D0$_3$) with different degrees of order  are formed for iron-rich region depending on temperature and composition. This fact makes Fe-Al attractive for study of ordering reactions since bcc structure is stable in the wide concentration and temperature range.}
In the concentration range up to $\approx19$~at.\%Al there is the $\alpha$-region with disordered bcc A2~($\alpha$-Fe) phase. Whereas the ordered phases based on B2 (FeAl) and D0$_3$~(Fe$_3$Al) structures form with a higher Al content~\cite{Knibloe1993,Ikeda2001,Leamy1967,Ershov2018}.
The D0$_3$ phase is stable below 825~K 
for compounds with $18 < x < 37$~at.\%Al while
the B2 phase occurs in concentration range between 23 and 54~at.\% Al depending on temperature (see Fig.~\ref{fig_0}). 
In general, there are two "order-disorder" (B2$\rightarrow$A2 and D0$_3\rightarrow$B2) and one paramagnetic (PM)$\rightarrow$ferromagnetic (FM) (A$2^{\mathrm{PM}}\rightarrow$A$2^{\mathrm{FM}}$) transformation for Fe-Al compounds~\cite{Ikeda2001, Stein}. Both transformations (B2$\rightarrow$A2 and D0$_3\rightarrow$B2) are the second order ones, B2$\rightarrow$A2 observes for compounds with $23 < x < 45$~at.\%Al whereas D0$_3\rightarrow$B2 takes place in the narrow concentration range and the maximal transition temperature of 818 K is observed at 26.5~at.\%.

{It should be noted that the Fe-Al phase diagram includes some features related to the areas of the coexistence of these three structures.
One of them is the small triangular area between 815 and 900~K for compositions with $21.3 < x < 22.8$at.\%Al where the FM disordered A2 phase coexists with the PM B2 phase~\cite{Ikeda2001, Stein, Okamoto}. 
The other feature is the two-phase (A2 and D0$_3$) region in the concentration range from 18 to 25~at.\% below 814~K. This region is characterized by the first-order "order-disorder" transition.}

{The tetragonal magnetostriction $\lambda_{001}$ of Fe-Al alloys has an unusual dependence as for Fe-rich Fe-Ga alloys~\cite{Clark_2008}.
Room temperature magnetostriction measurements of Fe-Al alloys indicated a five-fold rise in magnetostriction with Al content up to 30\% Al. Nevertheless, the $\lambda_{001}$ values for Fe-Al are smaller as compared to the Fe-Ga system. This finding can be explained by less magnitude of magnetoelastic constant $-b_1$ and the decreasing of shear modulus $C^{\prime}$ with Al content, which is not so large also as for Fe-Ga~\cite{Clark_2008}.
}

The aim of this paper is a complex study of structural and magnetic properties of Fe$_{100-x}$Al$_x$ ($5 \leq x \leq 25$~at.\%) alloys by the density functional theory at zero temperature with the finite-temperature Monte Carlo (MC) simulations.

\section{Methodology and calculation details}

{The \textit{ab initio} calculations were performed using the spin-polarized relativistic Korringa-Kohn-Rostoker code (SPR-KKR)~\cite{Ebert-sprkkr} implementing the KKR Green's function multiple scattering theory.
The exchange correlation energy was carried out in the Perdew–Burke–Ernzerhof (PBE)~\cite{Perdew-1996} generalized gradient approximation (GGA).}
{The single-site coherent potential approximation (CPA) successfully describing the properties of many compositionally disordered alloy systems was used to create the off-stoichiometric compositions.
An impurity of each atom type, Fe or Al, is placed in an effective CPA medium and then considers the alloy to be described by the weighted average of the two different impurity solutions.
As shown in~\cite{Banhart}, although the CPA deals with disorder systems, it can describe long-range order systems too.}

{The electronic structure calculations of Fe$_{100-x}$Al$_x$ ($5 \leq x \leq 25$~at.\%) were performed using the spin-polarized scalar-relativistic mode for three crystal cubic structures with different atomic order.
The fully ordered D0$_3$ structure with space group $Fm\bar{3}m$ (No. 225) was created by the unit cell consisting of four atoms.
The Al and Fe atoms were located at the $4a\ (0;\ 0;\ 0)$ Wyckoff position, the Fe atoms were occupied $4b\ (0.5;\ 0.5;\ 0.5)$ and $8c\ (0.25;\ 0.25;\ 0.25)$ Wyckoff positions. For stoichiometric Fe$_{75}$Al$_{25}$, $4a$ site was fully occupied by Al.
The partially ordered B2 phase with space group $Pm\bar{3}m$ (No. 221) was created using two atoms per unit cell, where Fe and Al atoms randomly were occupied the $1b\ (0.5;\ 0.5;\ 0.5)$ site while $1a\ (0;\ 0;\ 0)$ site was occupied by Fe atoms.  
For disordered A2 phase ($Im\bar{3}m$ space group, No. 229) Fe and Ga atoms were distributed at the $2a\ (0;\ 0;\ 0)$ Wyckoff position.
The calculations of optimized lattice constants were performed as single-point energy calculations at several different volumes for each structure under study. 
To evaluate equilibrium volume more precisely, the total energy curves as a function of volume were fitted by the  Birch–Murnaghan equation of state}.

{For the determination of tetragonal shear modulus $C^{\prime}$, the cubic structures were deformed from their optimized geometries along the $z$~axis assuming a volume-conserving mode $\varepsilon_x= \varepsilon_y = - 1/2 \varepsilon_z$ in the range of $\varepsilon = \pm2\%$.}
$C^{\prime}$ was calculated from the $\varepsilon_z$-dependent  total energy $E_\mathrm{tot.}$ according to the following equation~\cite{Wang2013},
\begin{equation}
    C^{\prime}=\frac{(C_{11}-C_{12})}{2}= \displaystyle\frac{1}{3V_0}\frac{d^2E_\mathrm{tot.}}{d\varepsilon^2}
    \label{shear modulus}
\end{equation}
here $V_0$ is the equilibrium volume of the unit cell.

For A2, B2, D0$_3$ cubic structures, magnetostrictive coefficient $\lambda_{001}$ was determined through the tetragonal dependence of magnetocrystalline anisotropy energy $E_\mathrm{{MCA}}$ as~\cite{Wang2013}
\begin{equation}
\label{eq_lambda}
\lambda_{001} = \frac{2}{3V_0 C^{\prime}}\frac{dE_{\mathrm{{MCA}}}}{d\varepsilon} =
-\frac{b_1}{3C^{\prime}},
\end{equation}
%%%%%%%%%%%%%%
here, $-b_1$  is magnetoelastic constant.
The $E_\mathrm{{MCA}}$ was calculated using the torque method implemented in the SPR-KKR package.

{The Heisenberg exchange parameters $J_{ij}$ were evaluated by way of KKR multiple scattering formalism proposed by Liechtenstein et al. ~\cite{Liechtenstein}.
The spin-polarized scalar-relativistic Dirac Hamiltonian mode and PBE approach for exchange-correlation potential were considered to calculate the $J_{ij}$ parameters as a function of distance between interacting atoms. 
Using the calculated $J_{ij}$ and Fe magnetic moments ($\mu^{\mathrm{Fe}}$) as input data, the finite-temperature MC simulations of Heisenberg model without magnetic field and anisotropy terms were done.}
\begin{equation}
    {\cal H}= - \sum_{ij}J_{ij}\mathbf{S}_i\mathbf{S}_j.
    \label{heis_ham}
\end{equation}
Here, $\mathbf{S}_{i}=\left(S_{i}^{x},\ S_{i}^{y},\ S_{i}^{z}\right)$ is a classical Heisenberg spin variable $\left|S_i\right|=1$.
{The $J_{ij}$ were taken into account up to the 6 coordinational spheres and can be ferromagnetic (FM, $J_{ij}>0$) or antiferromagnetic (AFM, $J_{ij}<0$).}
The modelling was performed on a 3925 atoms lattice with periodic boundary conditions using the Metropolis algorithm~\cite{LandauBinder}.
{The MC step at each site corresponded to an enumeration of randomly selected $N$ lattice sites, where $N$ is the number of atoms in the lattice.
For a given temperature, the number of MC steps at each site was $5\times10^5$.}
The magnetic order parameter $m$ and total magnetization $M$ are defined by the following way,
\begin{equation}
\begin{array}{ll}
    m^{\mathrm{Fe}}=\displaystyle\frac{1}{N^{\mathrm{Fe}}}\sum_{i}\sqrt{\left(S_i^x\right)^{2}+\left(S_i^y\right)^{2}+\left(S_i^z\right)^{2}},\\[1mm] M=3\mu^{\mathrm{Fe}}m^{\mathrm{Fe}},
    \end{array}
\end{equation}
Curie temperature ($T_C$) was estimated from temperature dependencies of magnetization by plotting $M^{1/\beta\left(T\right)}$ function that decreases almost linearly with increasing temperature.
The intersection of  {$M^{1/\beta}\left(T\right)$}  curve with the $T$ axis indicates $T_C$.  Here,
$\beta$ is the critical index, and it is equal to $0.3646$ for the three-dimensional Heisenberg model~\cite{Huang-1987}.

\section{Results and discussion}

In the first step of our calculations, we have done the geometric optimization of cubic crystal structures.
{The calculated ground-state energies summarized in Table~\ref{tab1} show that the D0$_3$ structure has the lowest energy for all considered compounds.
The energy difference between B2 and D0$_3$  structure ($\Delta E_0^{\mathrm{B}2-\mathrm{D}0_3}$) is smaller than the respective values for A2 and D0$_3$ ones ($\Delta E_0^{\mathrm{A}2-\mathrm{D}0_3}$). 
For both structures, the $\Delta E_0^{\mathrm{B}2-\mathrm{D}0_3}$ and $\Delta E_0^{\mathrm{A}2-\mathrm{D}0_3}$ are increasing functions of Al concentration and  they reach 65~meV/atom and 101~meV/atom for stoichiometric Fe$_{75}$Al$_{25}$, respectively.}
\begin{table}[htb!]
\begin{center}
\caption{The ground state energy $E_0$ (eV/atom) of Fe$_{100-x}$Al$_x$ (${5\leq x \leq 25}$~at.\%) with different cubic crystal structures, obtained via \textit{ab initio} calculations at zero temperature.}
\label{tab1}
\begin{tabular}{l|cc c c }
\hline
$x$, at.\% &D0$_3$  &B2$^{\mathrm{FM}}$ & {B2$^{\mathrm{NM}}$} & A2    \\
\hline
5 &	$-33219.647$&	$-33219.644$& $-33219.179$&	$-33219.642$ \\
10&	$-31818.911$&	$-31818.898$& $-31818.488$&	$-31818.893$\\
15&	$-30418.175$&	$-30418.148$& $-30417.800$&	$-30418.137$\\
18&	$-29577.735$&	$-29577.697$& $-29577.390$&	$-29577.681$\\
20&	$-29017.442$&	$-29017.397$& $-29017.117$&	$-29017.376$\\
21&	$-28737.296$&	$-28737.247$& $-28736.981$&	$-28737.223$\\
24&	$-27896.859$&	$-27896.797$& --&	$-27896.764$\\
25&	$-27616.713$&	$-27616.648$& $-27616.437$&	$-27616.612$\\
\hline
\end{tabular}
\end{center}
\end{table}
{According to the experimental data, the B2 structure is paramagnetic~\cite{Miyatani}.
However, \textit{ab initio} calculations generally predict a FM ground state for B2 phase~\cite{Lechermann,Rhee} as in our case (see Table 1).
There are two ways to explain this finding. 
One is the presence of defects in an experimental sample that have a strong influence on the magnetic state.
The other is the fact that the magnetic state depends strongly on the degree of chemical order~\cite{Lechermann}.
As it was shown by Mohn \textit{et al.}~\cite{Mohn} and Rhee \textit{et al.}~\cite{Rhee} 
the correct \textit{ab initio} prediction of magnetic state could be done by describing exchange and correlation within the local-density approximation including Hubbard parameters (LDA +$U$).}
{In the present work, we performed electronic structure calculations for the nonmagnetic state of the B2 phase (labeled B2$^{\mathrm{NM}}$) with an account of exchange-correlation energy in GGA approximation.
As can be seen from Table~\ref{tab1} the obtained ground state energies for the nonmagnetic state of B2  are higher than those for the disordered A2 structure.
In further discussion, all results for the B2 structure will be present in the FM state.}

\begin{figure}[!htb]
\centering
\includegraphics[height=6cm,clip]{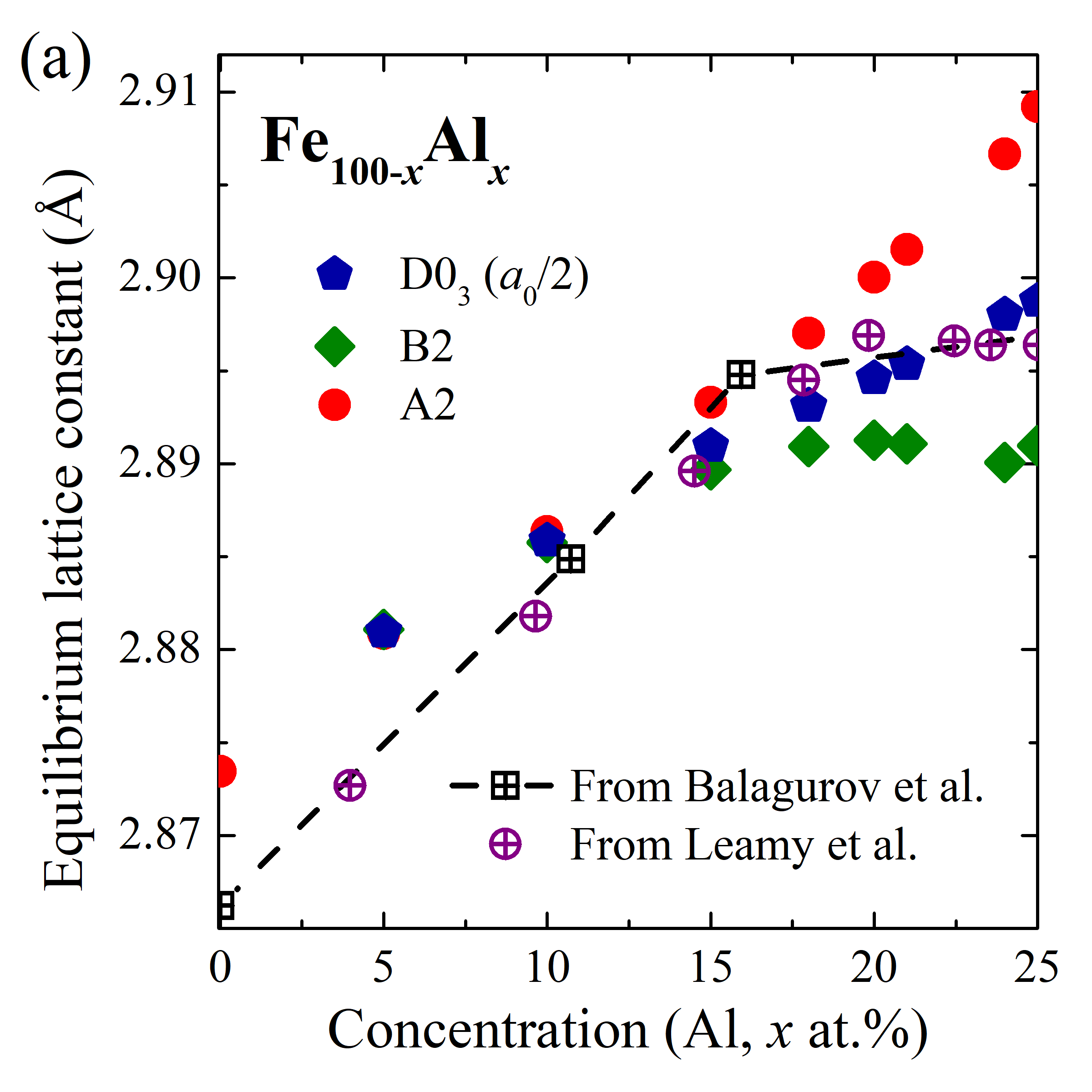}
\includegraphics[height=6cm,clip]{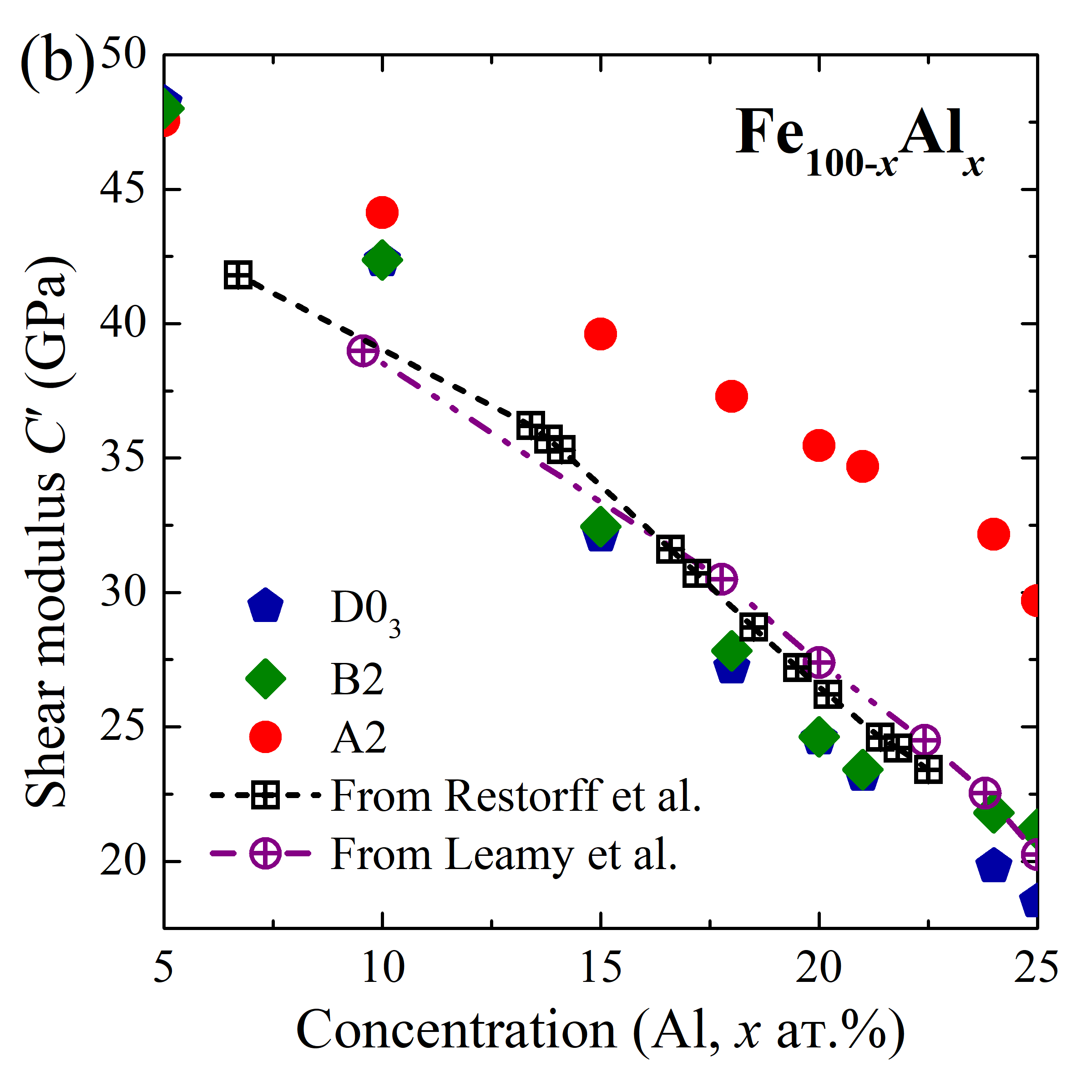}
\hfil
\caption{(Color online)~(a)~Calculated lattice parameters $a_0$ ($a_0/2$ for D0$_3$) and (b)~shear moduli as a function of the Al composition of Fe$_{100-x}$Al$_x$ (${5\leq x \leq 25}$~at.\%) alloys with A2, B2 and D0$_3$ structures.
The experimental results at room temperature were taken from Leamy~et~al.~\cite{Leamy1967}, Balagurov~et~al.~\cite{Balagurov}, and Restorff~et~al.~\cite{Restorff}. 
 }
\label{fig_1}
\end{figure}

Fig.~\ref{fig_1}~(a) shows the equilibrium lattice parameter $a_0$ ($a_0/2$ for D0$_3$) of phases corresponding to a minimum value of energy $E_0$ as a function of Al concentration for D0$_3$, B2, and A2 structures of Fe$_{100-x}$Al$_x$ alloys in comparison with available experimental data taken from~\cite{Leamy1967,Balagurov}.
For all cubic structures calculated herein, the lattice parameter is found to increase with increasing Al content.
The differences between $a_0^{\mathrm{A}2}-a_0^{\mathrm{D}0_3}$ and $a_0^{\mathrm{D}0_3}-a_0^{\mathrm{B}2}$ increase too and numerically are quite similar.
At the 18~at.\% of Al concentration, data for A2 and D0$_3$ are equally distant from the dash experimental line.
With further increasing of Al content, the slope of the theoretical curve $a_0^{\mathrm{A}2}(x)$ becomes higher than the experimental one.
The $a_0^{\mathrm{D}0_3}(x)$ curve slope is similar to the experiment that corresponds to neutron diffraction experiments by Balagurov et al.~\cite{Balagurov}.
According to~\cite{Balagurov}, the D0$_3$ phase corresponds to the main structural phase in the range from 21 to 34~at.\%.

%---Shear modulus and lambda------------
The concentration dependencies of calculated shear moduli  of D0$_3$, B2, and A2 structures are shown in Fig.~\ref{fig_1}~(b).
For comparison, experimental data taken from works~\cite{Leamy1967,Restorff} are plotted on the graph.
The $C^{\prime}$ decreases linearly with increasing Al content, and magnitudes of shear modulus for considered structures are close to each other for Al content up to 10 at.\%.
The anomaly in the $C^{\prime}$ behavior 
takes place for compounds with $x > 15$~at.\% Al. It could be related to the structural transformations which occur in the alloys as indicated by the atomic volume changes (See Fig.~\ref{fig_1}~(a)). 
Thus a difference in the $C^{\prime}$ behavior between D0$_3$(B2) and A2 has been attributed to the transition from disordered to ordered phase~\cite{Cullen}.

\begin{figure}[!htb]
\centering
\includegraphics[height=6cm,clip]{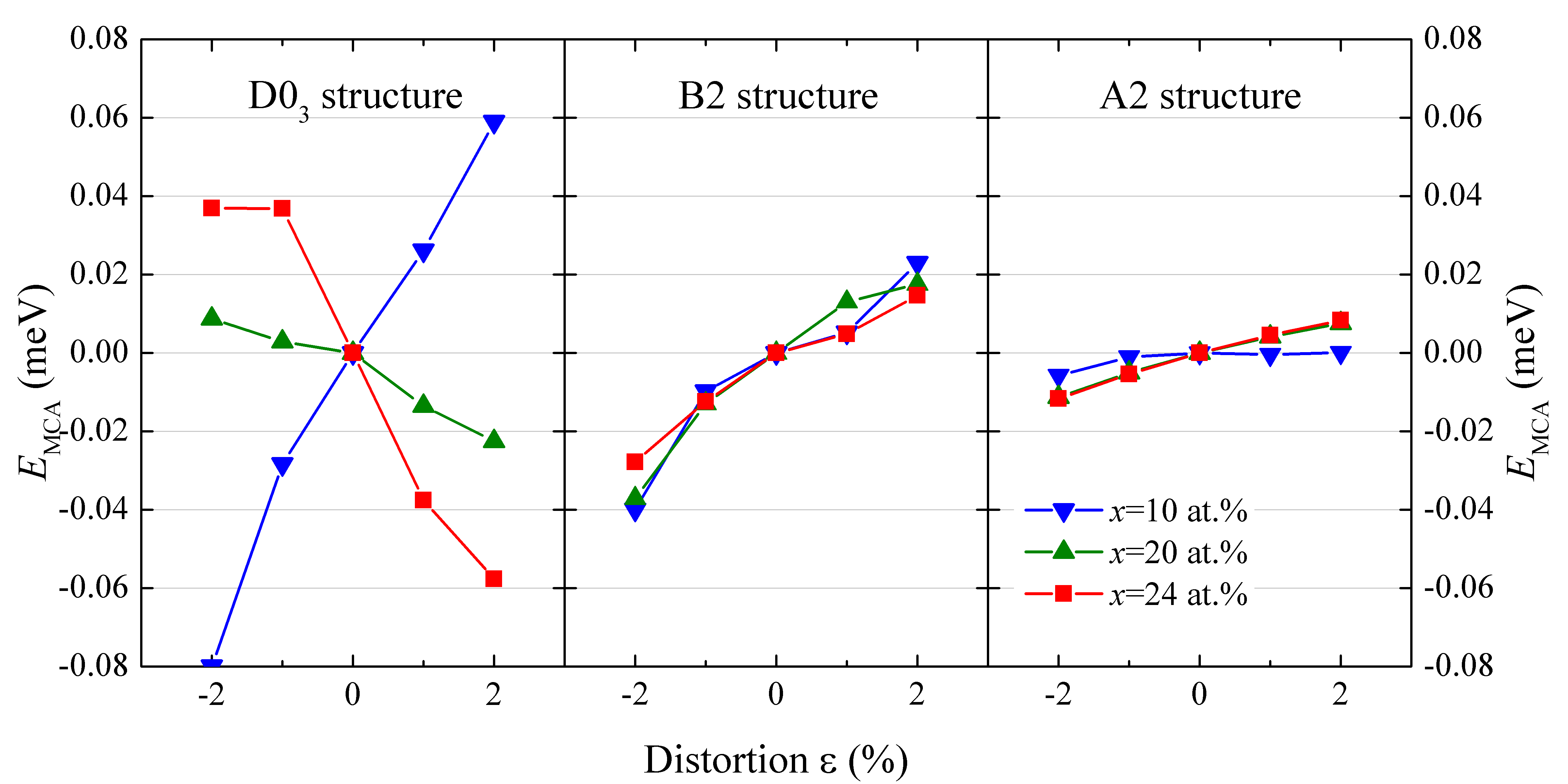}
\hfil
\caption{(Color online)~Calculated strain dependencies of magnetocrystalline anisotropy energy $E_{\mathrm{MCA}}$ of Fe$_{100-x}$Al$_x$ ($x=10,\ 20,\ 24$~at.\%) with D0$_3$, B2, and A2 structures.}
\label{fig_2}
\end{figure}

{Let us consider next the results of the magnetocrystalline anisotropy energy calculations $E_{\mathrm{MCA}}$ for Fe$_{100-x}$Al$_x$ (${5\leq x \leq 25}$~at.\%) with A2, D0$_3$ and B2 structures using the torque method.  
To demonstrate the tendency of $E_{\mathrm{MCA}}(\varepsilon)$ behavior for all structures, in Fig.~\ref{fig_2} we show
the results for compounds with $x=10,\ 20,\ 24$~at.\%. 
For A2 and B2 structures, the strain dependencies of $E_{\mathrm{MCA}}(\varepsilon)$ demonstrate the linear behavior with a positive slope, which slightly increases for the A2 phase and it is nearly constant for the B2 phase with increasing Al content.
In the case of D0$_3$ structure, an increase in Al content 20~at.\% leads to linear decrease of $E_{\mathrm{MCA}}(\varepsilon)$  curve and the slope of $E_{\mathrm{MCA}}(\varepsilon)$ is changed from positive to negative for Fe$_{80}$Al$_{20}$.}

\begin{figure}[!htb]
\centering
\includegraphics[height=6cm,clip]{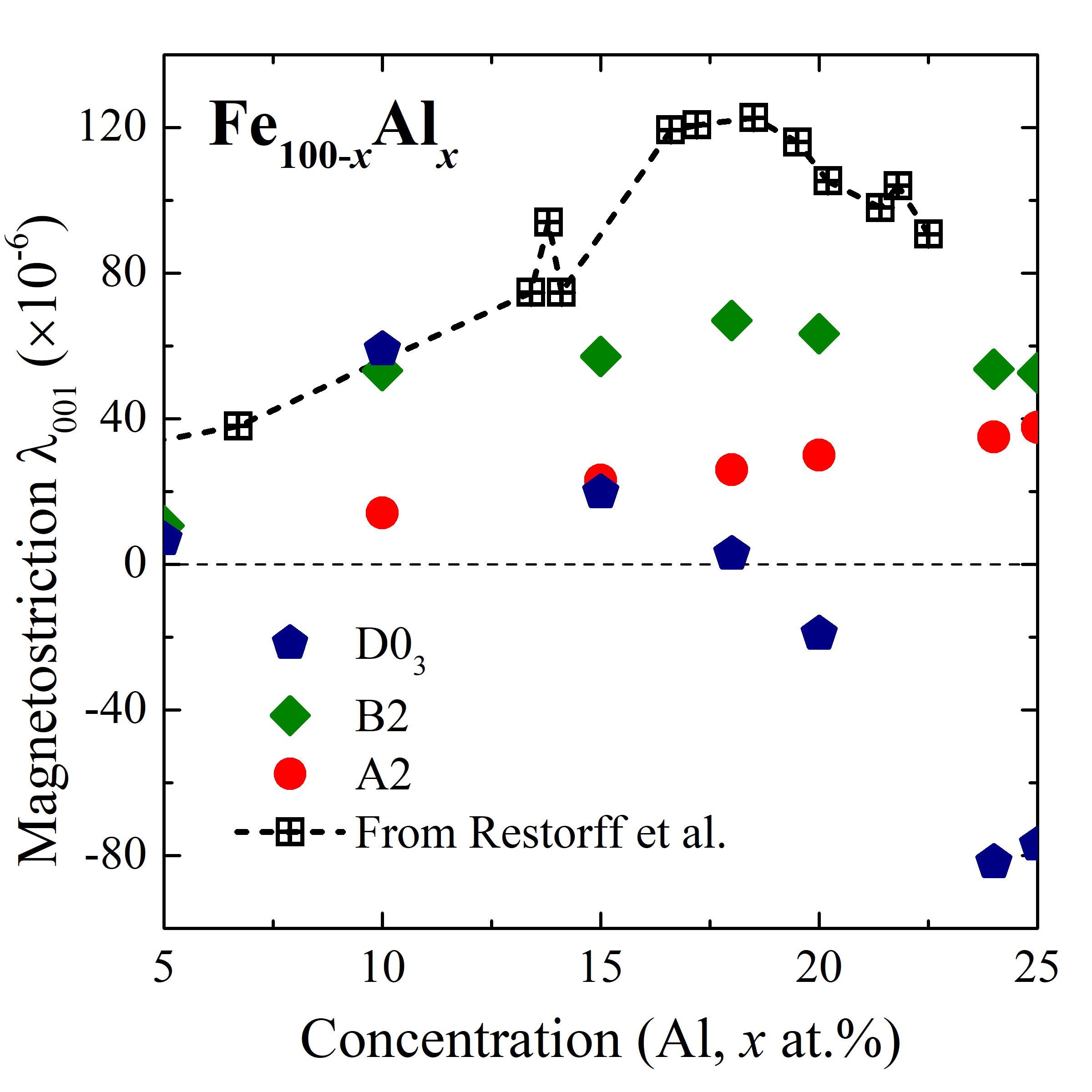}
\hfil
\caption{(Color online)~Calculated tetragonal magnetostriction constants of Fe$_{100-x}$Al$_x$ (${5\leq x \leq 25}$~at.\%) with D0$_3$, B2, and A2 structures. The experimental results were taken from Restorff~et~al.~\cite{Restorff}.}
\label{fig_3}
\end{figure}

The theoretical results of tetragonal magnetostriction $\lambda_{001}$ calculated by Eq.~(\ref{eq_lambda}) are presented in Fig.~\ref{fig_3}.
As can be seen, the A2 and B2 structures provide a positive contribution on tetragonal magnetostriction, while D0$_3$ has a positive magnitude only up to 18 at.\%.
{This tendency is dependent on $E_{\mathrm{MCA}}$ behavior.
The higher values of theoretical shear moduli $C^{\prime}$ could explain the numerically lesser value of $\lambda_{001}$ for the A2 structure compared with the B2 phase.}
A similar investigation of tetragonal magnetostriction using the torque method was done for Fe-Ga alloys~\cite{Matyunina}.
It has been shown that the main contribution to tetragonal magnetostriction in the region of Ga concentrations $21\leq x \leq 25$~at.\% makes the B2 phase.
{One of the earliest models proposed to explain the giant magnetostriction of Fe-Ga alloys was the question of Ga 2-nd nearest-neighbor (B2-like) pairing~\cite{Buschow-2012}. 
In the D0$_3-$like ordered phase, this pairing is strongly suppressed.}

\begin{figure}[!htb]
\centering
\includegraphics[height=6cm,clip]{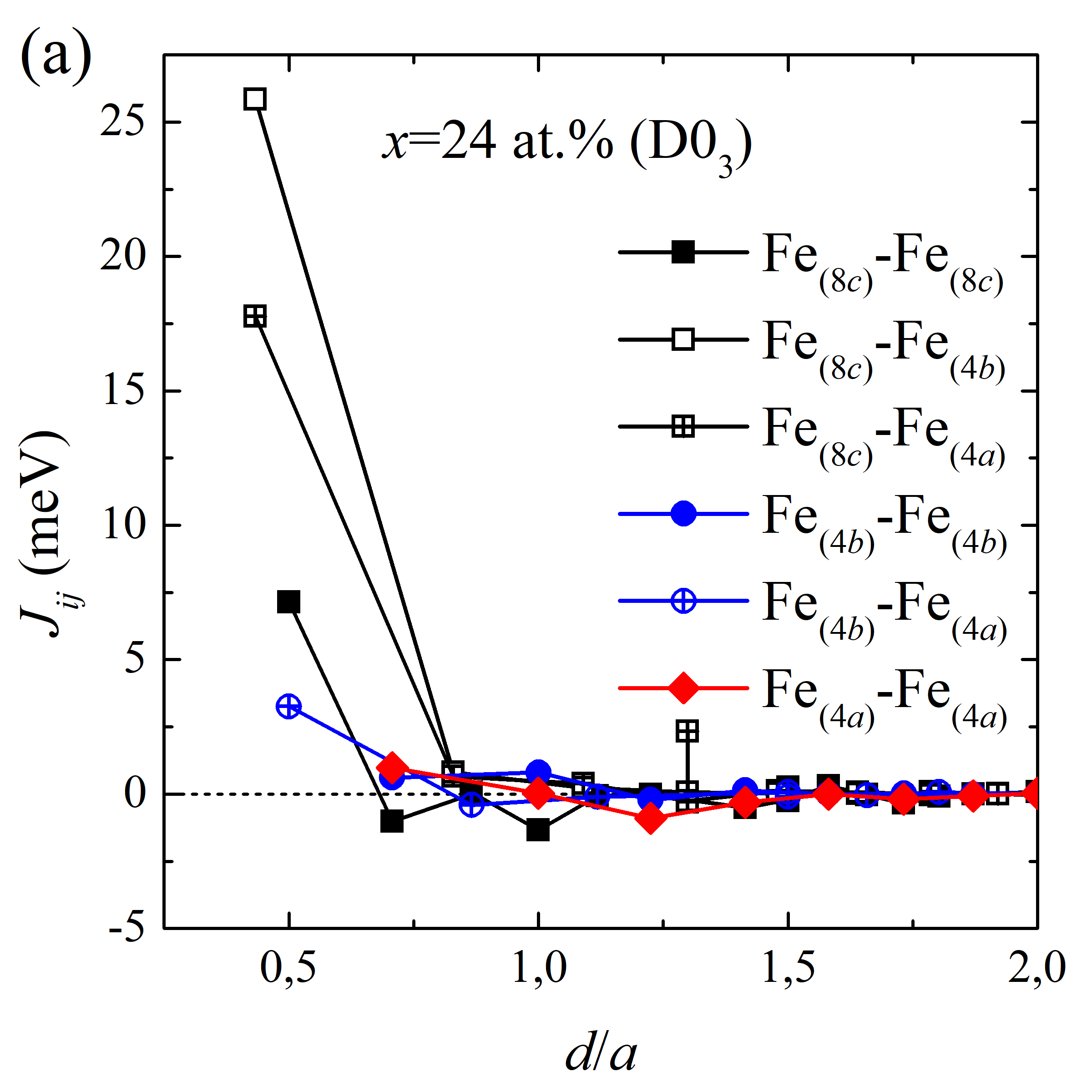}
\includegraphics[height=6cm,clip]{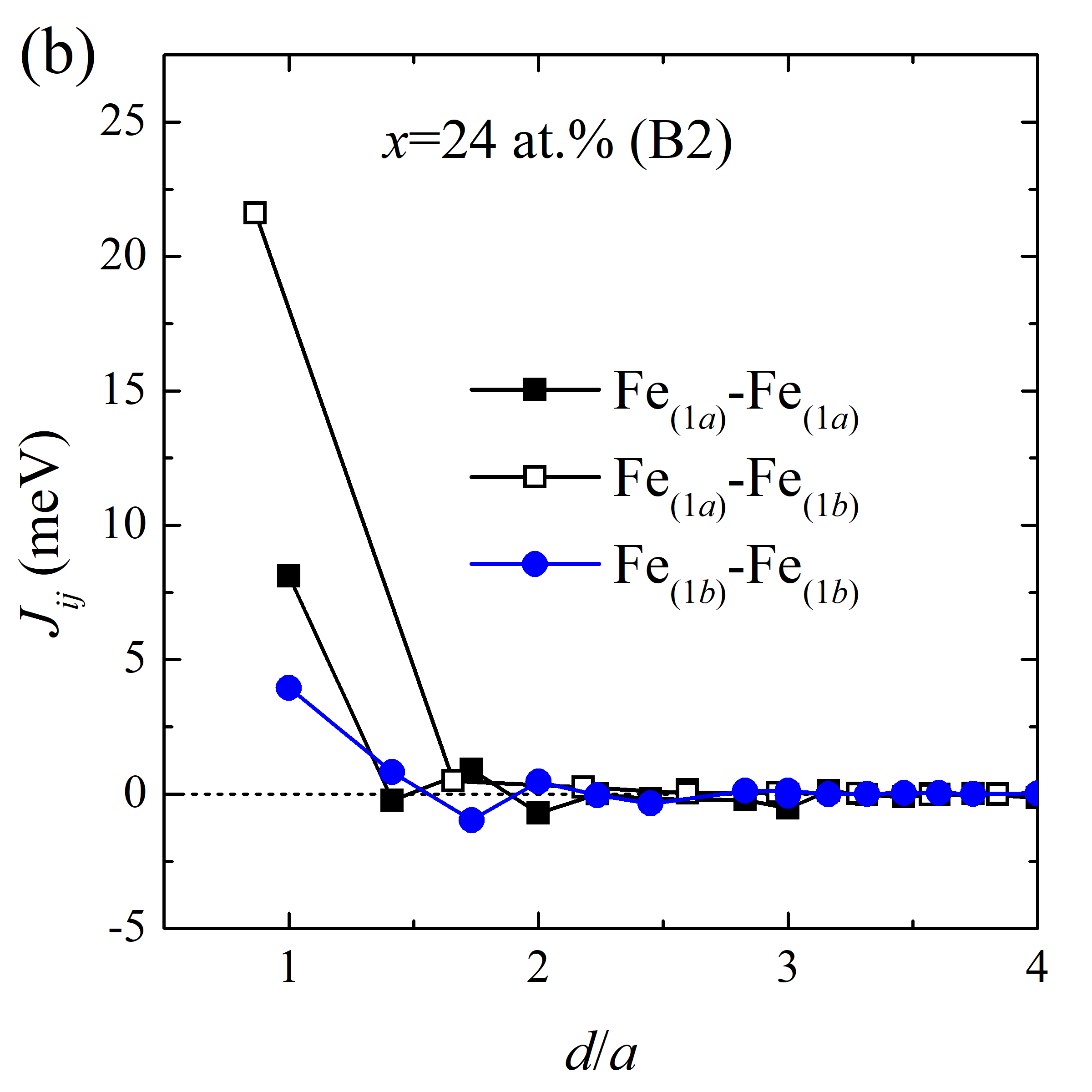}
\includegraphics[height=6cm,clip]{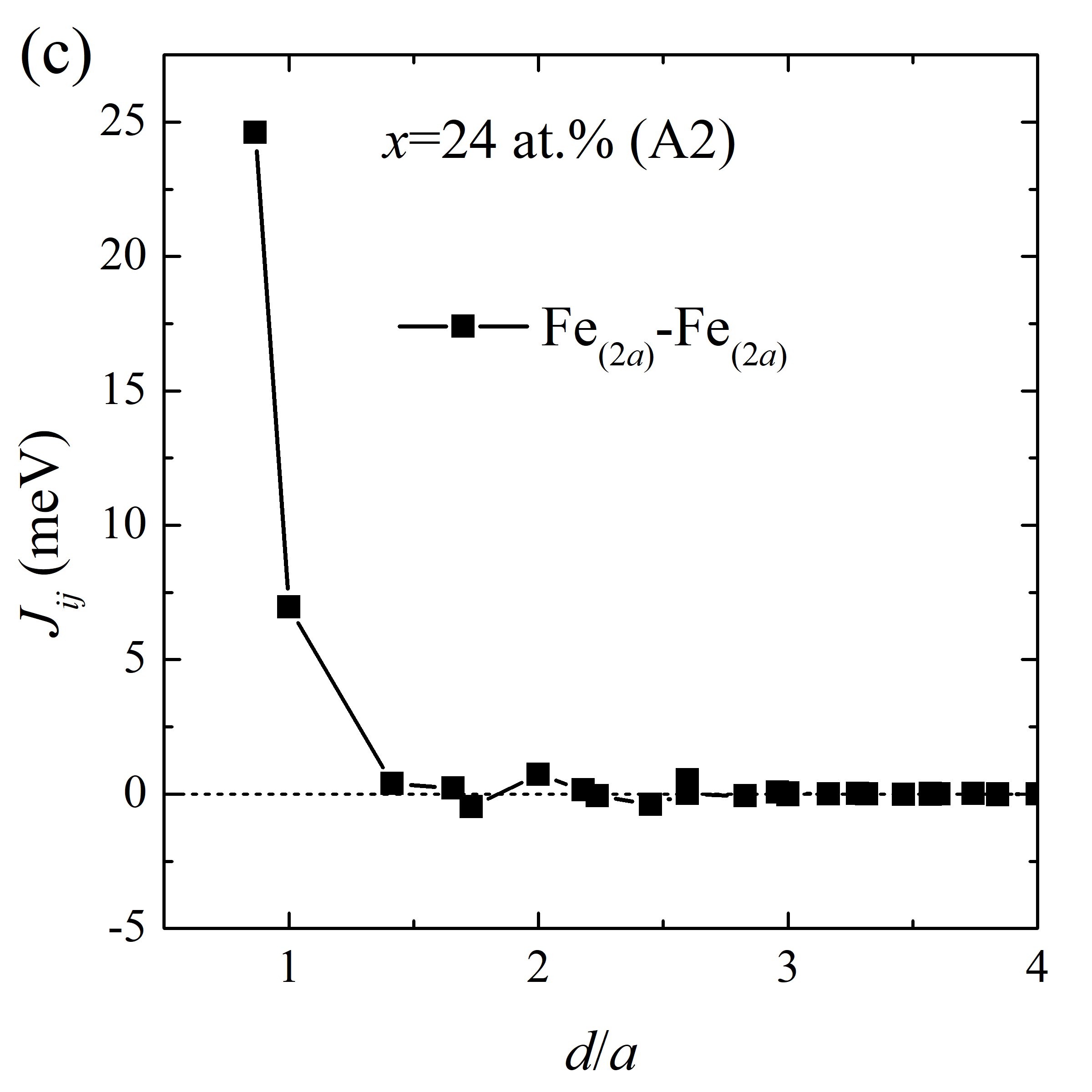}
\includegraphics[height=6cm,clip]{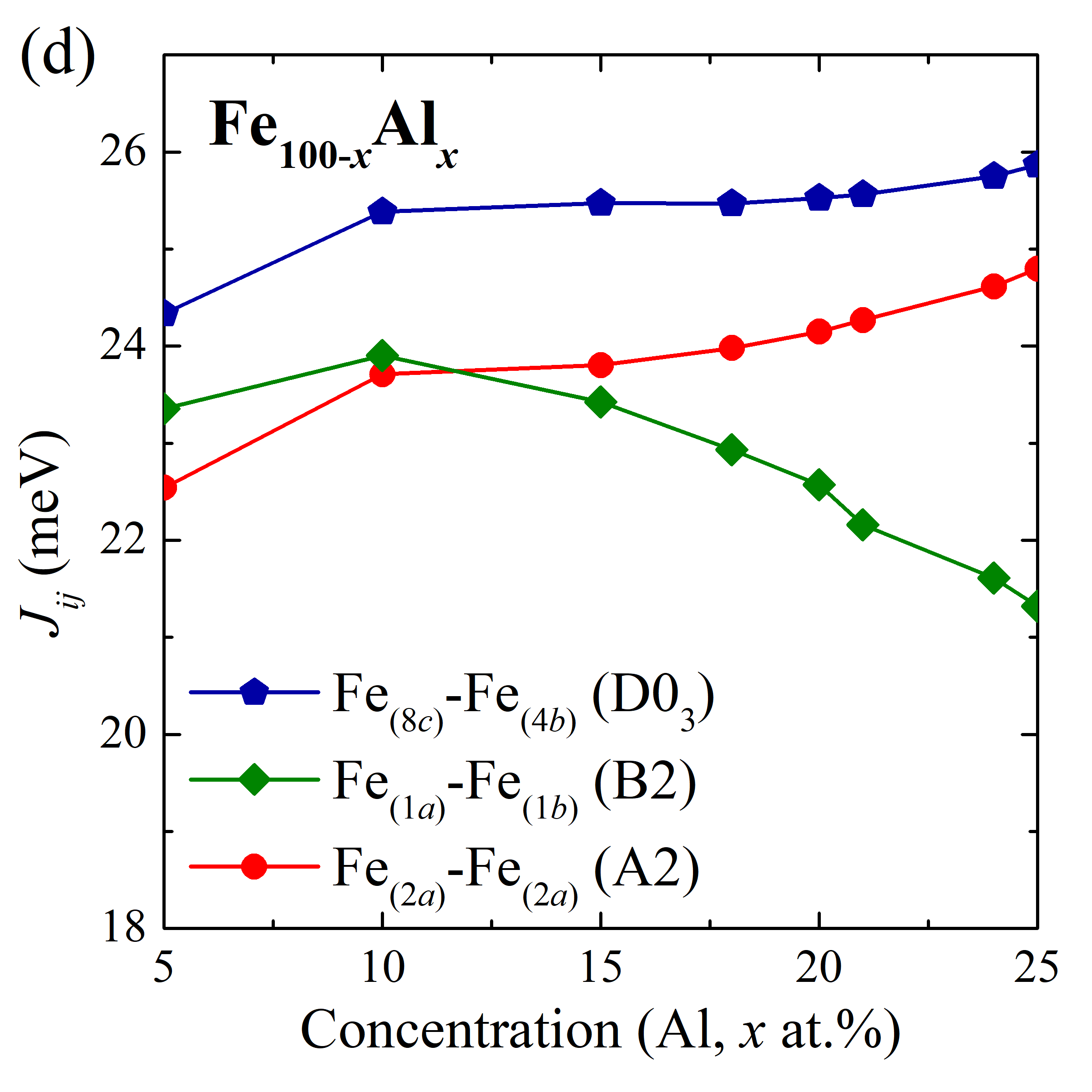}
\hfil
\caption{(Color online)~Calculated exchange coupling parameters $J_{ij}$ as a function of distance ($d/a$) between atoms $i$ and $j$ for (a)~D0$_3$, (b)~B2, and (c)~A2 structures of Fe$_{76}$Al$_{24}$ alloy.~(d)~Calculated $J_{ij}$ between first neighbors of iron atoms as a function of Al concentration of Fe$_{100-x}$Al$_x$ (${5\leq x \leq 25}$). The subscripts of Fe atoms correspond to Wyckoff positions.}
\label{fig_4}
\end{figure}

{The magnetic exchange coupling parameters $J_{ij}$ were calculated using the equilibrium lattice parameters and the SPR-KKR package by employing CPA.
We considered the FM state of D0$_3$, B2 and A2 structures of Fe$_{100-x}$Al$_x$ (${5\leq x \leq 25}$).
In general, the $J_{ij}$ shows damping oscillatory behavior for all compositions and structures under-considered.
The largest FM-interaction observes between the first nearest neighbour of Fe-Fe atoms.
Fig.~\ref{fig_4}(a)--(c) show the exchange as a function of distance $d/a$ between atoms $i$ and $j$ for D0$_3$, B2, and A2 structures in composition with 24~at.\% of Al content. 
Calculated $J_{ij}$ between first neighbors of iron atoms as a function of Al concentration is depicted in Fig.~\ref{fig_4}~(d).}
{As we mentioned in Section 2, in the fully ordered D0$_3$ structure, the Fe atoms are located on the $4a$, $4b$, and $8c$ Wyckoff positions.
In the B2 phase, Fe atoms occupy $1b$ and $1a$ Wyckoff positions, and in A2, there is one type of iron on the $2a$ Wyckoff position.
According to these locations, in fig.~\ref{fig_4} all Fe atoms are marked with the subscripts.
For the D0$_3$ and B2 structures, the intersublattice interactions Fe$_{4b}$-Fe$_{8c}$ and Fe$_{1b}$-Fe$_{1a}$ provide the largest contribution to the exchange due to shorter distances compared to the intrasublattice interactions.
As can be seen in fig.~\ref{fig_4}~(d), for the D0$_3$ and A2 structures,  the behavior of $J_{ij}\left(x\right)$ curves are similar to each other and the largest values of $J_{ij}$ are obtained in composition Fe$_{75}$Al$_{25}$. 
In the case of the B2 structure, the $J_{ij}$ is a decreasing function of Al content.}

The constants of magnetic exchange interactions and magnetic moments obtained from \textit{ab initio} calculations were used as input parameters to simulate the temperature dependences of magnetization and to estimate the Curie temperatures using Monte Carlo simulation.
The calculated Curie temperatures for D0$_3$, B2 and A2 structures are listed in 
Table~\ref{tab2}. 
\begin{table}[htb!]
\begin{center}
\caption{Curie temperatures $T_C$ (K) of Fe$_{100-x}$Al$_x$ (${5\leq x \leq 25}$) with D0$_3$, B2, and A2 structures, obtained via MC calculations. For comparison experimental data from~\cite{Stein} are shown. }\label{tab2}
\begin{tabular}{l|cc c|cc c| cc }
\hline
\multirow{2}{*}{$x$, at.\%}& \multicolumn{3}{c|}{MC calculations ($T_C$)}& \multicolumn{2}{c}{Experiment ($T_C$)}\\
 & D0$_3$ & B2 & A2 & D0$_3$ & A2 \\
\hline
5 & 907  &	1235&	1213 & $-$ & 1035\\
10&	1159 &	1350&	1258 & $-$ & 1018\\
15&	1300 &	1360&	1238 & $-$ & 991\\
18&	1328 &	1369&	1214 & $-$ & 967\\
20&	1338 &	1349&	1193 & $-$ & 937\\
21&	1345 &	1334&	1175 & $-$ & $-$\\
24&	1250 &	1218&	1131 & 781 & $-$\\
25&	1074 &	1206&	1127 & 758 &$-$\\
\hline
\end{tabular}
\end{center}
\end{table}
%
%The Curie temperature for all structures under consideration continuously decreases from 10~at.\% of Al with increasing Al content.
%From experimantal results it is follow
%Opposite to experimental data, the calculated Curie temperatures for A2 phase  increase from 10 at.\% for A2 and B2 phases. Curie temperatures for D0$_3$ structure increase up to 20 at.\% and then decrease. 
%MZ
It follows from experimental results, for the A2 structure, Curie temperatures continuously decrease with increasing Al concentration. In opposite to experimental results, calculated Curie temperature dependence for A2 structure has a maximum value at 10 at.\%.
Curie temperatures for B2 and D0$_3$ phases also increase up to 18 (for B2) and 20 (for D0$_3$)~at.\% and then decrease.
It should be noted calculated Curie temperatures have overestimated values in comparison to experimental data.
%MZ
This tendency could be explained by increasing magnetic exchange interaction parameters $J_{ij}$ (See Figure~\ref{fig_4}).

%\begin{figure}[!htb]
%\centering
%\includegraphics[height=6cm,clip]{Fig_5.png}
%\hfil
%\caption{(Color online)~Curie temperatures $T_C$ (K) of Fe$_{100-x}$Al$_x$ (${5\leq x \leq 25}$) with D0$_3$, B2, and A2 structures, obtained via MC calculations. 
%Experimental data is from~\cite{Stein}}
%\label{fig_5}
%\end{figure}

\section{Conclusion}

The complex investigation of structural and magnetic properties of Fe$_{100-x}$Al$_x$ alloys ($5 \leq x \leq 25$~at.\%) was done by means of \textit{ab initio} methods and Monte Carlo simulations. 
Experimentally observed crystal structures D0$_3$, B2, and A2 were considered. 
{Conducted theoretical calculations of the ground state energies showed that all considered phases are stable, while D0$_3$ is energetically favourable for the whole Al concentration range.}
It was found that the calculated equilibrium lattice constants increased with Al content.
In the range from 20 to 25 at.\%, the theoretical data of $a_0$ for the D0$_3$ phase are closer to the experiment that corresponds to the existence of D0$_3$ as the main structural phase.
Shear moduli $C^{\prime}$ of A2, B2 and D0$_3$ structures are linearly decreased with adding Al atoms in compositions.
The significant difference between D0$_3$ (B2) and A2 values of $C^{\prime}$ in an area higher than 15~at.\% Al content could be related to the atomic volume changes.

In A2 and B2 structures, the tetragonal magnetostriction constants are positive, while in D0$_3$, this parameter is positive only up to 18 at.\%.
This tendency depends on magnetocrystalline anisotropy energy behaviour as a function of Al content.
It was shown that the obtained values of $\lambda_{001}$ for the B2 structure are closer to the experimental one.
This result could be explained by the proposed model of formation B2-like short-range order that provide the giant magnetostriction in Fe-based alloys.

Using exchange interaction constants as input parameters, the Curie temperatures were estimated with the help of Monte Carlo simulations. 
The calculated Curie temperature dependence for the A2 structure has a maximum value at 10 at.\% while from experimental results, for the A2 structure, Curie temperatures continuously decrease with increasing of Al concentration.

\section*{Acknowledgment}

This work was supported by the Ministry of Science and Higher Education of the Russian Federation within the framework of the Russian State Assignment under contract No.~{075-00992-21-00}. MVM gratefully acknowledges the Advanced science research foundation of the Chelyabinsk State University.

\end{document}